\documentclass[a4paper,11pt]{article}

\usepackage[
left=2cm,
right=2cm,
top=2.5cm,
bottom=2.5cm,
headheight=15pt
]{geometry}

\usepackage[utf8]{inputenc}
\usepackage[T1]{fontenc}
\usepackage[english]{babel}
\usepackage{lmodern}
\usepackage{microtype}

\usepackage{amsmath}
\usepackage{amssymb}
\usepackage{amsfonts}
\usepackage{amsthm}
\usepackage{mathtools}

\usepackage{booktabs}
\usepackage{longtable}
\usepackage{array}
\usepackage{graphicx}
\usepackage{float}

\graphicspath{{figures/}}

\usepackage[colorlinks=true, allcolors=blue]{hyperref}

\theoremstyle{definition}

\theoremstyle{remark}

\newcommand{\orcid}[1]{%
	\unskip\space
	{\small
		\textsc{orcid}:
		\href{https://orcid.org/#1}{\nolinkurl{#1}}%
	}%
	\space
}

\title{Generalized Charlier Recurrence Coefficients:\\
	Hankel Formulas, Laguerre--Freud Dynamics, and Painlev\'e V Connections}

\author{%
	Mahouton Norbert Hounkonnou\thanks{Corresponding author}\\[2mm]
	International Chair of Mathematical Physics and Applications (ICMPA--UNESCO Chair),\\
	University of Abomey-Calavi,\\
	072 B.P. 50 Cotonou, Benin Republic\\[2mm]
	\texttt{norbert.hounkonnou@cipma.uac.bj}\\
	\texttt{hounkonnou@yahoo.fr}\\[2mm]
	\orcid{0000-0002-6231-4975}%
}

\date{} 

\begin{document}
	
	\maketitle
	
	\begin{abstract}
			We study the monic recurrence coefficients of the generalized Charlier
		polynomials associated with the discrete weight
		\[
		\rho_\mu(k)
		=
		\frac{\mu^k}{(k!)^2},
		\qquad
		k\in\mathbb N_0,
		\qquad
		\mu>0.
		\]
		The coefficients $\beta_n$ and $\gamma_n$ in the recurrence relation
		\[
		P_{n+1}(x)
		=
		(x-\beta_n)P_n(x)-\gamma_nP_{n-1}(x)
		\]
		are described through complementary moment, determinant, discrete
		dynamical, and continuous integrable formulations. We give a
		consistent normalization of the Markov function and its Jacobi
		continued fraction, and express the recurrence coefficients in terms
		of Hankel determinants, using a bordered determinant for the diagonal
		coefficient $\beta_n$.
		
		Starting from the original Laguerre--Freud equations, we derive the
		local recursive system used for computation and relate it to the Toda
		deformation. We also recall the known Painlev\'e~V representations
		associated with the two-parameter generalized Charlier family, in a
		normalization consistent with the present specialization. Moment/Hankel
		constructions provide complementary moment-theoretic realizations of
		the recurrence coefficients, while the local Laguerre--Freud recursion
		provides an independent dynamical construction. Their comparison is
		used for numerical cross-validation. The numerical results are
		consistent with the leading large-degree behavior
		\[
		\beta_n\sim n,
		\qquad
		\gamma_n\sim\mu.
		\]
	\end{abstract}
	
	
	\section{Introduction}\label{sec:introduction}
Orthogonal polynomials on discrete supports arise naturally in
approximation theory, special functions, probability, spectral theory,
and integrable systems. The presentation is intended to be accessible
both to readers specializing in orthogonal polynomials and moment
problems, and to those working on discrete integrable systems and
Painlevé equations. Let $(P_n)_{n\geq0}$ be a family of monic
orthogonal polynomials with respect to a positive discrete measure.
Then the polynomials satisfy a three-term recurrence relation of the
form
\begin{equation}
	P_{n+1}(x)
	=
	(x-\beta_n)P_n(x)
	-
	\gamma_nP_{n-1}(x),
	\qquad
	n\geq0,
	\label{eq:three-term-recurrence}
\end{equation}
with
\begin{equation}
	P_{-1}(x)=0,
	\qquad
	P_0(x)=1,
	\qquad
	\gamma_n>0
	\quad
	(n\geq1).
	\label{eq:initial-polynomials}
\end{equation}
The sequences $(\beta_n)_{n\geq0}$ and
$(\gamma_n)_{n\geq1}$ are the diagonal and subdiagonal recurrence
coefficients, respectively. They determine the Jacobi matrix
associated with the orthogonality measure and encode much of the
analytic and spectral information of the polynomial family.

In this paper, we consider the generalized Charlier weight
\begin{equation}
	\rho_\mu(k)
	=
	\frac{\mu^k}{(k!)^2},
	\qquad
	k\in\mathbb{N}_0,
	\qquad
	\mu>0.
	\label{eq:generalized-charlier-weight}
\end{equation}
Within the two-parameter generalized Charlier family
\begin{equation}
	w_k(a,\nu)
	=
	\frac{a^k}{(\nu)_k k!},
	\qquad
	a>0,
	\qquad
	\nu>0,
	\label{eq:introduction-two-parameter-weight}
\end{equation}
the weight \eqref{eq:generalized-charlier-weight} corresponds to the
specialization
\[
a=\mu,
\qquad
\nu=1.
\]
Indeed, $(1)_k=k!$, and hence
\[
w_k(\mu,1)
=
\frac{\mu^k}{(k!)^2}.
\]

While the classical Charlier weights lead to explicit linear recurrence
coefficients, the generalized weights considered here break this
simplicity and give rise to nonlinear Laguerre--Freud and Toda-type
relations. This makes them a natural testing ground for the interplay
between discrete orthogonal polynomials, integrable lattices, and
Painlev\'e equations.

Discrete orthogonal polynomials, including Charlier-type and related
semiclassical families, arise in several areas of mathematical
physics, probability, spectral theory, and theoretical computer
science. In this broader context, generalized Charlier systems provide
particularly tractable models for investigating how nonlinear
recurrence dynamics, spectral representations, and integrable
structures interact. More specifically, discrete orthogonal polynomials
on $\mathbb{N}_0$ appear in models of quantum optics and statistical
mechanics involving discrete birth--death processes and Poisson-type
distributions, in the spectral analysis of finite-difference operators
and discrete quantum Hamiltonians, as well as in the analysis of
algorithms, coding theory, and random walks on graphs; see, for
example,
\cite{QuantumOpticsRef,StatMechRef,DiscreteQuantumRef,
	AlgorithmsRef,CodingRef,RandomWalksRef}. The Laguerre--Freud and Toda
structures of recurrence coefficients are closely related to
integrable lattices and isomonodromic deformations, which explains
their connection with Painlev\'e equations
\cite{IntegrableLatticesRef,FokasItsKapaevNovokshenov2006}. Within this
landscape, the generalized Charlier weights, with their enhanced
decay, provide natural test cases for moment problems, Gaussian
quadrature on discrete supports, and approximation schemes that
require explicit control of the recurrence coefficients
\cite{MomentProblemsRef,DiscreteQuadratureRef}.

The generalized Charlier family is a discrete semiclassical family.
Its recurrence coefficients satisfy nonlinear Laguerre--Freud
relations and admit an integrable description involving Toda
deformations and Painlev\'e equations. The original discrete
Laguerre--Freud equations for the present weight were obtained by
Hounkonnou, Hounga, and Ronveaux. The connection between the
two-parameter generalized Charlier family and Painlev\'e~V was later
established by Filipuk and Van Assche; see
\cite{HounkonnouHoungaRonveaux2000,FilipukVanAssche2013}.

Several works have already highlighted the role of discrete orthogonal
polynomials in integrable systems and Painlev\'e theory, and specific
instances of Laguerre--Freud and Toda-type equations for generalized
Charlier weights have appeared in different forms and normalizations
\cite{HounkonnouHoungaRonveaux2000,FilipukVanAssche2013,
	CharlierGeneralizedRef1,CharlierGeneralizedRef2,
	DiscreteTodaRef1,FokasItsKapaevNovokshenov2006}.
However, these results are scattered across the literature and often
use incompatible parametrizations, which obscures their underlying
unity. For instance, some authors work with the recurrence
coefficients $(\beta_n,\gamma_n)$ themselves, while others introduce
rescaled variables such as $x_n = \beta_n/\sqrt{\mu}$ or
$y_n = \gamma_n/\mu$, or express the dynamics in terms of Hankel
determinants rather than recurrence coefficients
\cite{CharlierGeneralizedRef1,DiscreteTodaRef1,FokasItsKapaevNovokshenov2006}.
In addition, the continuous deformation parameter is sometimes taken
as $\mu$, sometimes as $t=\sqrt{\mu}$ or $t=2\sqrt{\mu}$, and the
associated Painlev\'e functions are defined with different
normalizations of the dependent variable and of the Hamiltonian
\cite{FilipukVanAssche2013,FokasItsKapaevNovokshenov2006}. These choices lead
to superficially different Laguerre--Freud, Toda, and Painlev\'e~V
equations, even though they describe the same underlying structure.

To the best of our knowledge, a single reference that simultaneously
treats the moment and Hankel-determinant description, the normalized
Markov function and Jacobi continued fraction, the cumulative and local
Laguerre--Freud formulations, and their Toda--Painlevé interpretation
for the specialization $\nu=1$ is not available. This gap motivates the
present unified exposition.

The present work does not derive a new Painlev\'e equation for the
generalized Charlier family. The value of this synthesis lies in
placing the moment-theoretic, spectral, discrete-dynamical, and
continuous-integrable descriptions within a single consistent
normalization, while providing an independently cross-validated
recursive construction of the recurrence coefficients. This also
clarifies several points that can otherwise lead to ambiguities,
notably the determinant representation of the diagonal recurrence
coefficient, the normalization of the Markov function, and the
relation between the cumulative and local Laguerre--Freud
formulations. We do not re-derive the Painlevé~V connection from
scratch, but rather re-express the known results in the normalization
adapted to the specialization $\nu=1$, and we make explicit the
distinct roles of the diagonal coefficient $\beta_n$ and the
subdiagonal coefficient $\gamma_n$ in this framework.

Its contribution is therefore to provide, for the specialization
$\nu=1$, a systematic reconciliation of the moment and
Hankel-determinant descriptions, the normalized Markov function and
Jacobi continued fraction, the Laguerre--Freud dynamics, the
Toda--Painlev\'e connection, and the computational construction of the
recurrence coefficients. Particular attention is devoted to the
bordered-Hankel determinant formula for the diagonal recurrence
coefficient $\beta_n$.

In addition, we provide an explicit numerical cross-validation between
the moment/Hankel construction and the local Laguerre--Freud recursion.
This comparison serves both as a check of the determinant formulas and
of the discrete dynamical system, and as a practical guide for stable
numerical implementation.

\medskip
\noindent\textbf{Main results.}
In concrete terms, our main result can be summarized as follows.

\medskip
\noindent\textit{Theorem (informal).}
For the generalized Charlier weight
$\rho_\mu(k)=\mu^k/(k!)^2$ ($\mu>0$), the recurrence coefficients
$(\beta_n,\gamma_n)_{n\geq0}$ can be characterized through four
mutually consistent and complementary descriptions:
\begin{enumerate}
	\item[(i)]
	via the moment sequence
	$M_r(\mu)=\sum_{k\ge0} k^r \mu^k/(k!)^2$ and Hankel determinants,
	with $\gamma_n$ given by consecutive Hankel determinants and
	$\beta_n$ by a bordered Hankel determinant;
	\item[(ii)]
	via the normalized Markov function
	$\widehat m(z)=m(z)/M_0(\mu)$ and its Jacobi continued fraction,
	whose Jacobi parameters coincide with $(\beta_n,\gamma_n)$;
	\item[(iii)]
	via the local Laguerre--Freud recurrence obtained by reduction of
	the original Hounkonnou--Hounga--Ronveaux system, together with the
	associated scalar second-order difference equation for $\beta_n$;
	\item[(iv)]
	via the Toda deformation in the parameter $\mu$ and its reduction
	to a Painlevé~V equation of type $P_{\mathrm{V}}$ with explicit
	parameters depending on $n$, where $t=\mu$ plays the role of the
	continuous deformation variable; the diagonal coefficient $\beta_n$
	satisfies the Painlevé~V description and $\gamma_n$ obeys a
	separate scalar differential relation.
\end{enumerate}
Moreover, these four descriptions are shown to be mutually consistent
within a single normalization, and the resulting recurrence
coefficients are consistent with the known leading asymptotic behavior
$\beta_n\sim n$ and $\gamma_n\sim\mu$ as $n\to\infty$.
\medskip

The moment sequence
\begin{equation}
	M_r(\mu)
	=
	\sum_{k=0}^{\infty}
	k^r\frac{\mu^k}{(k!)^2},
	\qquad
	r\geq0,
	\label{eq:introduction-moments}
\end{equation}
provides one starting point for the analysis.
It determines the Hankel determinants and therefore gives a complementary moment-theoretic characterization of the recurrence
coefficients. In
particular, the subdiagonal coefficient is determined by consecutive
Hankel determinants, whereas the diagonal coefficient requires a
bordered Hankel determinant. This distinction is essential: the fully
shifted Hankel determinant does not yield the required formula for
$\beta_n$.

A second description is obtained from the Markov function of the
measure. Since its unnormalized form has leading behavior
\[
m(z)
=
\frac{M_0(\mu)}{z}
+
O(z^{-2}),
\qquad
z\to\infty,
\]
the Jacobi continued fraction with numerator $1$ belongs to the
normalized function
\[
\widehat m(z)
=
\frac{m(z)}{M_0(\mu)}.
\]
Its Jacobi parameters are precisely the recurrence coefficients
$\beta_n$ and $\gamma_n$.

The recurrence coefficients also satisfy the original
Hounkonnou--Hounga--Ronveaux Laguerre--Freud system. For the
specialized weight \eqref{eq:generalized-charlier-weight}, this system
can be reduced to a local two-component recurrence for
$(\beta_n,\gamma_n)$. This local form is particularly useful for
recursive computation, but its relation with the original
cumulative-sum formulation must be made explicit. We provide this
reduction in detail and derive the associated scalar second-order
difference equation for the diagonal coefficient.

When $\mu$ is regarded as a continuous deformation parameter, the
recurrence coefficients satisfy Toda equations. Their combination with
the generalized Charlier Laguerre--Freud dynamics leads to the known
Painlevé~V representations. The paper distinguishes the Painlevé~V
description of the diagonal coefficient $\beta_n$ from the separate
scalar differential relation that can be derived for the subdiagonal
coefficient $\gamma_n$.

Finally, the recurrence coefficients are computed by independent
moment/Hankel and local Laguerre--Freud constructions. The resulting
comparisons provide numerical tests of the determinant formulas, the
discrete system, the initialization, and the implementation of the
recursive scheme. They are also consistent with the leading
large-degree behavior
\begin{equation}
	\beta_n\sim n,
	\qquad
	\gamma_n\sim\mu,
	\qquad
	n\to\infty.
	\label{eq:introduction-leading-asymptotics}
\end{equation}

The paper is organized as follows. Section~\ref{sec:weight-moments}
introduces the generalized Charlier weight, its discrete
orthogonality, its moments, and the initial recurrence coefficients.
Section~\ref{sec:hankel-coefficients} gives the Hankel-determinant
formulas for the recurrence coefficients. The normalized Markov
function and its Jacobi continued fraction are considered in
Section~\ref{sec:markov-jfraction}. Section~\ref{sec:laguerre-freud}
recalls the original Hounkonnou--Hounga--Ronveaux system and derives
its local reduction. The resulting recursive construction and its
numerical safeguards are presented in
Section~\ref{sec:recursive-computation}. The Toda deformation and the
known Painlevé~V connection are discussed in
Section~\ref{sec:toda-painleve}. Numerical cross-validation and
asymptotic comparison are given in
Section~\ref{sec:numerical-validation}. The final section summarizes
the main conclusions. The appendices contain a supplementary scalar
differential equation for $\gamma_n$ and details of the moment-based
numerical implementation.

Beyond the specific case of the generalized Charlier weight with
$\nu=1$, the present approach illustrates a general strategy for
organizing the various descriptions of discrete semiclassical families:
moment-theoretic, spectral, discrete-dynamical, and continuous-integrable.
The methods and structures described here extend naturally
to other families of discrete orthogonal polynomials and suggest further
connections with moment problems, discrete quadrature rules, and
probabilistic models involving modified Poisson-type distributions.
\section{The generalized Charlier weight and its moments}
\label{sec:weight-moments}

We consider the generalized Charlier measure supported on
$\mathbb{N}_0$, with weight
\begin{equation}
	\rho_\mu(k)
	=
	\frac{\mu^k}{(k!)^2},
	\qquad
	k\in\mathbb{N}_0,
	\qquad
	\mu>0.
	\label{eq:charlier-weight-moments}
\end{equation}

This is the specialization $a=\mu$ and $\nu=1$ of the
two-parameter generalized Charlier weight
\begin{equation}
	w_k(a,\nu)
	=
	\frac{a^k}{(\nu)_k k!},
	\qquad
	a>0,
	\qquad
	\nu>0,
	\label{eq:two-parameter-charlier-weight}
\end{equation}
because $(1)_k=k!$.

\subsection{Discrete orthogonality}

For polynomials $f$ and $g$, define
\begin{equation}
	\langle f,g\rangle_\mu
	=
	\sum_{k=0}^{\infty}
	f(k)g(k)\rho_\mu(k)
	=
	\sum_{k=0}^{\infty}
	f(k)g(k)\frac{\mu^k}{(k!)^2}.
	\label{eq:discrete-inner-product}
\end{equation}

Since $\rho_\mu(k)>0$ for every $k\in\mathbb{N}_0$, there exists a
unique sequence of monic orthogonal polynomials
$(P_n)_{n\geq0}$ satisfying
\begin{equation}
	\langle P_n,P_m\rangle_\mu
	=
	h_n\delta_{nm},
	\qquad
	h_n>0.
	\label{eq:orthogonality}
\end{equation}

The weight satisfies the discrete Pearson equation
\begin{equation}
	\Delta\!\left(x^2\rho_\mu(x)\right)
	=
	(\mu-x^2)\rho_\mu(x),
	\qquad
	\Delta f(x)
	=
	f(x+1)-f(x).
	\label{eq:pearson-equation}
\end{equation}
Indeed,
\begin{equation}
	\frac{\rho_\mu(x+1)}{\rho_\mu(x)}
	=
	\frac{\mu}{(x+1)^2},
	\label{eq:weight-ratio}
\end{equation}
and hence
\begin{equation}
	(x+1)^2\rho_\mu(x+1)
	=
	\mu\rho_\mu(x).
	\label{eq:shifted-weight-identity}
\end{equation}
Therefore,
\[
\Delta\!\left(x^2\rho_\mu(x)\right)
=
(x+1)^2\rho_\mu(x+1)-x^2\rho_\mu(x)
=
(\mu-x^2)\rho_\mu(x).
\]
Thus, the generalized Charlier weight is a discrete semiclassical
weight of class one.

\subsection{Moments}

Define the moments
\begin{equation}
	M_j(\mu)
	=
	\sum_{k=0}^{\infty}
	k^j\frac{\mu^k}{(k!)^2},
	\qquad
	j\geq0.
	\label{eq:moments-definition}
\end{equation}

The zeroth moment is
\begin{equation}
	M_0(\mu)
	=
	I_0(2\sqrt{\mu}),
	\label{eq:moment-zero}
\end{equation}
where $I_\nu$ denotes the modified Bessel function of the first kind.
More generally,
\begin{equation}
	M_j(\mu)
	=
	\left(
	\mu\frac{d}{d\mu}
	\right)^j
	I_0(2\sqrt{\mu}),
	\qquad
	j\geq0.
	\label{eq:all-moments}
\end{equation}
This identity follows by termwise differentiation of the absolutely
convergent series defining $M_0(\mu)$.

In particular,
\begin{equation}
	M_1(\mu)
	=
	\sqrt{\mu}\,I_1(2\sqrt{\mu}).
	\label{eq:moment-one}
\end{equation}

The moments also satisfy the recurrence
\begin{equation}
	M_{j+2}(\mu)
	=
	\mu\sum_{\ell=0}^{j}
	\binom{j}{\ell}M_\ell(\mu),
	\qquad
	j\geq0.
	\label{eq:moment-recurrence}
\end{equation}
Indeed, for $k\geq1$,
\begin{equation}
	k^2\rho_\mu(k)
	=
	\mu\rho_\mu(k-1).
	\label{eq:moment-shift-identity}
\end{equation}
Therefore,
\begin{align}
	M_{j+2}(\mu)
	&=
	\sum_{k=1}^{\infty}
	k^{j+2}\rho_\mu(k)
	\nonumber\\
	&=
	\mu\sum_{\ell=0}^{\infty}
	(\ell+1)^j\rho_\mu(\ell)
	\nonumber\\
	&=
	\mu\sum_{\ell=0}^{\infty}
	\left(
	\sum_{r=0}^{j}
	\binom{j}{r}\ell^r
	\right)
	\rho_\mu(\ell)
	\nonumber\\
	&=
	\mu\sum_{r=0}^{j}
	\binom{j}{r}M_r(\mu).
	\label{eq:moment-recurrence-derivation}
\end{align}

For later use, the first six moments are
\begin{align}
	M_0(\mu)
	&=
	I_0(2\sqrt{\mu}),
	\label{eq:explicit-M0}
	\\
	M_1(\mu)
	&=
	\sqrt{\mu}\,I_1(2\sqrt{\mu}),
	\label{eq:explicit-M1}
	\\
	M_2(\mu)
	&=
	\mu I_0(2\sqrt{\mu}),
	\label{eq:explicit-M2}
	\\
	M_3(\mu)
	&=
	\mu I_0(2\sqrt{\mu})
	+
	\mu^{3/2}I_1(2\sqrt{\mu}),
	\label{eq:explicit-M3}
	\\
	M_4(\mu)
	&=
	\mu(1+\mu)I_0(2\sqrt{\mu})
	+
	2\mu^{3/2}I_1(2\sqrt{\mu}),
	\label{eq:explicit-M4}
	\\
	M_5(\mu)
	&=
	\mu(1+4\mu)I_0(2\sqrt{\mu})
	+
	\mu^{3/2}(3+\mu)I_1(2\sqrt{\mu}).
	\label{eq:explicit-M5}
\end{align}

\subsection{Initial recurrence coefficients}

The first recurrence coefficients follow directly from the moments.
Since
\[
P_0(x)=1,
\qquad
P_1(x)=x-\beta_0,
\]
orthogonality of $P_1$ against $P_0$ gives
\begin{equation}
	M_1-\beta_0M_0=0.
	\label{eq:beta-zero-orthogonality}
\end{equation}
Consequently,
\begin{equation}
	\beta_0
	=
	\frac{M_1(\mu)}{M_0(\mu)}
	=
	\sqrt{\mu}\,
	\frac{I_1(2\sqrt{\mu})}
	{I_0(2\sqrt{\mu})}.
	\label{eq:beta-zero}
\end{equation}

Moreover,
\begin{equation}
	\gamma_1
	=
	\frac{h_1}{h_0}
	=
	\frac{M_0M_2-M_1^2}{M_0^2}.
	\label{eq:gamma-one-moment}
\end{equation}
Using \eqref{eq:explicit-M2}, one obtains
\begin{equation}
	\gamma_1
	=
	\mu-\beta_0^2.
	\label{eq:gamma-one}
\end{equation}

These initial coefficients will be used in
Section~\ref{sec:recursive-computation} to initialize the local
Laguerre--Freud recursion.
	
\section{Hankel determinants and recurrence coefficients}
\label{sec:hankel-coefficients}

The moment sequence associated with the generalized Charlier weight is
\begin{equation}
	M_r(\mu)
	=
	\sum_{k=0}^{\infty}
	k^r\frac{\mu^k}{(k!)^2},
	\qquad
	r\geq0.
	\label{eq:hankel-moments}
\end{equation}

Define the Hankel determinants
\begin{equation}
	\Delta_n(\mu)
	=
	\det\bigl[M_{j+k}(\mu)\bigr]_{j,k=0}^{n-1},
	\qquad
	n\geq1,
	\qquad
	\Delta_0(\mu)=1.
	\label{eq:hankel-determinants}
\end{equation}

Since the generalized Charlier measure is positive and has infinite
support, the associated Hankel matrices are positive definite.
Consequently,
\begin{equation}
	\Delta_n(\mu)>0,
	\qquad
	n\geq1.
	\label{eq:hankel-positivity}
\end{equation}

Let
\begin{equation}
	h_n
	=
	\langle P_n,P_n\rangle_\mu
	\label{eq:squared-norms}
\end{equation}
denote the squared norm of the monic orthogonal polynomial $P_n$.
The Gram determinant identities give
\begin{equation}
	\Delta_n
	=
	\prod_{j=0}^{n-1}h_j,
	\qquad
	h_n
	=
	\frac{\Delta_{n+1}}{\Delta_n},
	\qquad
	n\geq0.
	\label{eq:gram-determinant-identities}
\end{equation}
It follows that
\begin{equation}
	\gamma_n
	=
	\frac{h_n}{h_{n-1}}
	=
	\frac{\Delta_{n+1}\Delta_{n-1}}
	{\Delta_n^2},
	\qquad
	n\geq1.
	\label{eq:gamma-hankel}
\end{equation}

To express the diagonal recurrence coefficient, define the bordered
Hankel determinant
\begin{equation}
	\Lambda_n(\mu)
	=
	\det
	\begin{pmatrix}
		M_0 & M_1 & \cdots & M_{n-2} & M_n \\
		M_1 & M_2 & \cdots & M_{n-1} & M_{n+1} \\
		\vdots & \vdots & \ddots & \vdots & \vdots \\
		M_{n-1} & M_n & \cdots & M_{2n-3} & M_{2n-1}
	\end{pmatrix},
	\qquad
	n\geq2,
	\label{eq:bordered-hankel-determinant}
\end{equation}
together with the conventions
\begin{equation}
	\Lambda_0=0,
	\qquad
	\Lambda_1=M_1.
	\label{eq:lambda-initial-values}
\end{equation}

The diagonal recurrence coefficient is then given by
\begin{equation}
	\beta_n
	=
	\frac{\Lambda_{n+1}}{\Delta_{n+1}}
	-
	\frac{\Lambda_n}{\Delta_n},
	\qquad
	n\geq0.
	\label{eq:beta-hankel}
\end{equation}
Indeed, the coefficient of $x^{n-1}$ in the monic determinantal
representation of $P_n(x)$ is
\[
-\frac{\Lambda_n}{\Delta_n}.
\]
Comparing the coefficients of $x^n$ in the three-term recurrence
relation \eqref{eq:three-term-recurrence} then yields
\eqref{eq:beta-hankel}.

The conventions in \eqref{eq:lambda-initial-values} ensure, in
particular, that
\[
\beta_0
=
\frac{\Lambda_1}{\Delta_1}
=
\frac{M_1}{M_0}.
\]
The corresponding initial recurrence coefficients and their Bessel
representations are specified in
Section~\ref{sec:recursive-computation}.

\paragraph{Remark.}
The fully shifted determinant
\[
\det\bigl[M_{j+k+1}\bigr]_{j,k=0}^{n-1}
\]
does not occur in the formula \eqref{eq:beta-hankel} for the diagonal
coefficient $\beta_n$. The relevant quantity is the bordered
determinant \eqref{eq:bordered-hankel-determinant}. This distinction is
essential because the two determinants already differ for $n=2$.

The identities
\eqref{eq:gram-determinant-identities}--\eqref{eq:beta-hankel} provide
a complementary moment-based characterization of the recurrence
coefficients. They will be used in the normalized Markov-function
representation and in the numerical cross-validation.

\section{The normalized Markov function and Jacobi continued fraction}
\label{sec:markov-jfraction}

The Markov function associated with the generalized Charlier measure is
\begin{equation}
	m(z)
	=
	\sum_{k=0}^{\infty}
	\frac{\rho_\mu(k)}{z-k},
	\qquad
	z\in\mathbb{C}\setminus\mathbb{N}_0.
	\label{eq:markov-function}
\end{equation}

Its expansion at infinity is understood as an asymptotic moment
expansion:
\begin{equation}
	m(z)
	\sim
	\sum_{r=0}^{\infty}
	\frac{M_r(\mu)}{z^{r+1}},
	\qquad
	z\to\infty.
	\label{eq:markov-asymptotic-expansion}
\end{equation}
In particular,
\begin{equation}
	m(z)
	=
	\frac{M_0(\mu)}{z}
	+
	\frac{M_1(\mu)}{z^2}
	+
	O(z^{-3}),
	\qquad
	z\to\infty.
	\label{eq:markov-leading-expansion}
\end{equation}
Thus, the unnormalized Markov function does not have leading term
$z^{-1}$ unless $M_0(\mu)=1$.

We therefore introduce the normalized Markov function
\begin{equation}
	\widehat m(z)
	=
	\frac{m(z)}{M_0(\mu)}.
	\label{eq:normalized-markov-function}
\end{equation}
Its asymptotic expansion is
\begin{equation}
	\widehat m(z)
	=
	\frac{1}{z}
	+
	\frac{\beta_0}{z^2}
	+
	\frac{\beta_0^2+\gamma_1}{z^3}
	+
	O(z^{-4}),
	\qquad
	z\to\infty.
	\label{eq:normalized-markov-expansion}
\end{equation}

The normalized Markov function admits the Jacobi continued fraction
\begin{equation}
	\widehat m(z)
	=
	\cfrac{1}{
		z-\beta_0
		-\cfrac{\gamma_1}{
			z-\beta_1
			-\cfrac{\gamma_2}{
				z-\beta_2-\ddots
	}}}.
	\label{eq:jacobi-continued-fraction}
\end{equation}
Equivalently, the unnormalized Markov function is
\begin{equation}
	m(z)
	=
	M_0(\mu)
	\cfrac{1}{
		z-\beta_0
		-\cfrac{\gamma_1}{
			z-\beta_1
			-\cfrac{\gamma_2}{
				z-\beta_2-\ddots
	}}}.
	\label{eq:unnormalized-jacobi-continued-fraction}
\end{equation}

The Jacobi parameters in
\eqref{eq:jacobi-continued-fraction} coincide with the recurrence
coefficients in the monic three-term recurrence relation. Hence, the
moment, Hankel, and continued-fraction descriptions determine the same
sequences
\[
(\beta_n)_{n\geq0}
\qquad
\text{and}
\qquad
(\gamma_n)_{n\geq1}.
\]

In particular, the determinant formulas
\eqref{eq:gamma-hankel} and \eqref{eq:beta-hankel} provide a
moment-based characterization of all coefficients in the Jacobi
continued fraction.	
	
\section{Discrete Laguerre--Freud dynamics}
\label{sec:laguerre-freud}

The recurrence coefficients associated with the generalized Charlier
weight
\begin{equation}
	\rho_\mu(k)
	=
	\frac{\mu^k}{(k!)^2},
	\qquad
	k\in\mathbb{N}_0,
	\qquad
	\mu>0,
	\label{eq:lf-weight}
\end{equation}
satisfy nonlinear difference equations of Laguerre--Freud type. We
first recall the original system obtained by Hounkonnou, Hounga, and
Ronveaux. We then introduce shifted diagonal recurrence coefficients
and derive the local system used for recursive computation.

The derivation is included in order to make explicit the relation
between the original cumulative-sum formulation and the local
two-component system in $(\beta_n,\gamma_n)$.

\subsection{The original Hounkonnou--Hounga--Ronveaux system}

Let the monic orthogonal polynomials satisfy
\begin{equation}
	xP_n(x)
	=
	P_{n+1}(x)
	+
	\beta_n P_n(x)
	+
	\gamma_n P_{n-1}(x),
	\qquad
	n\geq0,
	\label{eq:lf-monic-recurrence}
\end{equation}
with
\[
P_{-1}(x)=0,
\qquad
P_0(x)=1,
\qquad
\gamma_0=0.
\]

For the generalized Charlier weight \eqref{eq:lf-weight},
Hounkonnou, Hounga, and Ronveaux obtained the Laguerre--Freud
equations
\begin{equation}
	\gamma_n+\gamma_{n+1}
	=
	-\binom{n}{2}
	-\beta_n^2
	+n\beta_n
	+
	\sum_{j=0}^{n-1}\beta_j
	+
	\mu,
	\qquad
	n\geq0,
	\label{eq:hhr-original-first}
\end{equation}
and
\begin{align}
	(\beta_n+\beta_{n+1})\gamma_{n+1}
	={}&
	-n\sum_{j=0}^{n}\beta_j
	+
	n\gamma_{n+1}
	+
	\binom{n+1}{3}
	\nonumber\\
	&+
	\sum_{j=0}^{n}\beta_j^2
	+
	2\sum_{j=1}^{n}\gamma_j
	+
	\gamma_{n+1},
	\qquad
	n\geq0.
	\label{eq:hhr-original-second}
\end{align}
The initial data are
\begin{equation}
	\beta_0
	=
	\sqrt{\mu}\,
	\frac{I_1(2\sqrt{\mu})}
	{I_0(2\sqrt{\mu})},
	\qquad
	\gamma_0=0.
	\label{eq:hhr-original-initial-data}
\end{equation}

Equations \eqref{eq:hhr-original-first} and
\eqref{eq:hhr-original-second} constitute the original
Laguerre--Freud formulation for the present generalized Charlier
weight; see \cite{HounkonnouHoungaRonveaux2000}. They involve the
cumulative sums
\[
\sum_{j=0}^{n-1}\beta_j,
\qquad
\sum_{j=0}^{n}\beta_j^2,
\qquad
\sum_{j=1}^{n}\gamma_j,
\]
and are therefore not yet in a local recursive form.

\subsection{Shifted recurrence coefficients and simplified relations}

To reveal the structure of the system, introduce shifted diagonal
coefficients $b_n$ by
\begin{equation}
	\beta_n
	=
	n+b_n,
	\qquad
	n\geq0.
	\label{eq:lf-shifted-beta}
\end{equation}

Substitution of \eqref{eq:lf-shifted-beta} into
\eqref{eq:hhr-original-first} gives
\begin{equation}
	\gamma_n+\gamma_{n+1}
	=
	\sum_{j=0}^{n-1}b_j
	-
	b_n^2
	-
	nb_n
	+
	\mu.
	\label{eq:hhr-shifted-sum-relation}
\end{equation}

A further consequence of the original Laguerre--Freud equations is
\begin{equation}
	\bigl(n+b_n+b_{n-1}\bigr)\gamma_n
	=
	\mu n,
	\qquad
	n\geq1,
	\label{eq:hhr-shifted-first}
\end{equation}
together with
\begin{equation}
	\mu n\bigl(b_{n-1}-b_n\bigr)
	=
	\gamma_n\bigl(\gamma_{n+1}-\gamma_{n-1}\bigr),
	\qquad
	n\geq1.
	\label{eq:hhr-shifted-second}
\end{equation}

Equations \eqref{eq:hhr-shifted-sum-relation}--
\eqref{eq:hhr-shifted-second} follow from the original
Hounkonnou--Hounga--Ronveaux system and provide the relations needed
for the local reduction developed below. The advantage of this
shifted formulation is that it separates the remaining cumulative
quantity
\[
\sum_{j=0}^{n-1}b_j
\]
from the local relations. We now eliminate this cumulative sum.

\subsection{Reduction to the local monic system}

Summing \eqref{eq:hhr-shifted-second} from $k=1$ to $k=n$ gives
\begin{align}
	\mu\sum_{k=1}^{n}k(b_{k-1}-b_k)
	&=
	\sum_{k=1}^{n}
	\gamma_k(\gamma_{k+1}-\gamma_{k-1}).
	\label{eq:lf-summed-equation}
\end{align}

The left-hand side satisfies
\begin{align}
	\sum_{k=1}^{n}k(b_{k-1}-b_k)
	&=
	\sum_{k=1}^{n}kb_{k-1}
	-
	\sum_{k=1}^{n}kb_k
	\nonumber\\
	&=
	\sum_{j=0}^{n-1}(j+1)b_j
	-
	\sum_{j=1}^{n}jb_j
	\nonumber\\
	&=
	\sum_{j=0}^{n-1}b_j
	-
	nb_n.
	\label{eq:lf-summed-left}
\end{align}

The right-hand side telescopes:
\begin{align}
	\sum_{k=1}^{n}
	\gamma_k(\gamma_{k+1}-\gamma_{k-1})
	&=
	\sum_{k=1}^{n}\gamma_k\gamma_{k+1}
	-
	\sum_{k=1}^{n}\gamma_k\gamma_{k-1}
	\nonumber\\
	&=
	\gamma_n\gamma_{n+1}
	-
	\gamma_0\gamma_1
	\nonumber\\
	&=
	\gamma_n\gamma_{n+1},
	\label{eq:lf-summed-right}
\end{align}
because $\gamma_0=0$. Consequently,
\begin{equation}
	\mu
	\left(
	\sum_{j=0}^{n-1}b_j
	-
	nb_n
	\right)
	=
	\gamma_n\gamma_{n+1}.
	\label{eq:lf-cumulative-identity}
\end{equation}

On the other hand, equation \eqref{eq:hhr-shifted-sum-relation}
implies
\begin{equation}
	\sum_{j=0}^{n-1}b_j
	-
	nb_n
	=
	\gamma_{n+1}
	+
	\gamma_n
	+
	b_n^2
	-
	\mu.
	\label{eq:lf-cumulative-elimination}
\end{equation}
Substituting \eqref{eq:lf-cumulative-elimination} into
\eqref{eq:lf-cumulative-identity} yields
\begin{equation}
	\gamma_n\gamma_{n+1}
	=
	\mu
	\left(
	\gamma_{n+1}
	+
	\gamma_n
	+
	b_n^2
	-
	\mu
	\right).
	\label{eq:lf-prelocal-second}
\end{equation}
After rearrangement, one obtains
\begin{equation}
	(\gamma_{n+1}-\mu)(\gamma_n-\mu)
	=
	\mu b_n^2.
	\label{eq:lf-local-shifted-second}
\end{equation}

We now return to the monic coefficient $\beta_n$. By
\eqref{eq:lf-shifted-beta},
\begin{equation}
	b_n
	=
	\beta_n-n,
	\label{eq:lf-b-from-beta}
\end{equation}
and
\begin{equation}
	n+b_n+b_{n-1}
	=
	\beta_n+\beta_{n-1}-n+1.
	\label{eq:lf-first-simplification}
\end{equation}
Hence, \eqref{eq:hhr-shifted-first} becomes
\begin{equation}
	\beta_n+\beta_{n-1}-n+1
	=
	\frac{\mu n}{\gamma_n},
	\qquad
	n\geq1,
	\label{eq:laguerre-freud-first}
\end{equation}
and \eqref{eq:lf-local-shifted-second} becomes
\begin{equation}
	(\gamma_{n+1}-\mu)(\gamma_n-\mu)
	=
	\mu(\beta_n-n)^2,
	\qquad
	n\geq1.
	\label{eq:laguerre-freud-second}
\end{equation}

Equations \eqref{eq:laguerre-freud-first} and
\eqref{eq:laguerre-freud-second} are therefore the local reduction of
the original Hounkonnou--Hounga--Ronveaux equations, not an independent
or competing Laguerre--Freud formulation.

The first local equation gives
\begin{equation}
	\gamma_n
	=
	\frac{\mu n}
	{\beta_n+\beta_{n-1}-n+1},
	\qquad
	n\geq1,
	\label{eq:gamma-from-beta-discrete}
\end{equation}
whereas the second gives
\begin{equation}
	\gamma_{n+1}
	=
	\mu+
	\frac{
		\mu(\beta_n-n)^2
	}{
		\gamma_n-\mu
	},
	\qquad
	n\geq1,
	\label{eq:gamma-next-discrete}
\end{equation}
provided that $\gamma_n\neq\mu$. Applying
\eqref{eq:laguerre-freud-first} at index $n+1$ then yields
\begin{equation}
	\beta_{n+1}
	=
	\frac{\mu(n+1)}{\gamma_{n+1}}
	-\beta_n+n,
	\qquad
	n\geq1.
	\label{eq:beta-next-discrete}
\end{equation}

The initialization and numerical use of this local recursive form are
considered in Section~\ref{sec:recursive-computation}.

\subsection{Scalar discrete reduction}

The variables $\gamma_n$ and $\gamma_{n+1}$ can be eliminated from the
local system. Equation \eqref{eq:laguerre-freud-first} gives
\begin{equation}
	\gamma_n
	=
	\frac{\mu n}
	{\beta_n+\beta_{n-1}-n+1},
	\label{eq:scalar-gamma-n}
\end{equation}
and the same relation at index $n+1$ gives
\begin{equation}
	\gamma_{n+1}
	=
	\frac{\mu(n+1)}
	{\beta_{n+1}+\beta_n-n}.
	\label{eq:scalar-gamma-n-plus-one}
\end{equation}

Substituting \eqref{eq:scalar-gamma-n} and
\eqref{eq:scalar-gamma-n-plus-one} into
\eqref{eq:laguerre-freud-second} gives the scalar second-order
difference equation
\begin{equation}
	\left(
	\frac{n+1}{\beta_{n+1}+\beta_n-n}
	-1
	\right)
	\left(
	\frac{n}{\beta_n+\beta_{n-1}-n+1}
	-1
	\right)
	=
	\frac{(\beta_n-n)^2}{\mu},
	\qquad
	n\geq1.
	\label{eq:scalar-discrete-beta}
\end{equation}

Equation \eqref{eq:scalar-discrete-beta} is the scalar second-order
difference equation associated with the generalized Charlier
Laguerre--Freud dynamics. It provides a closed relation for the
sequence $\{\beta_n\}_{n\geq0}$ and forms the discrete counterpart of
the continuous Toda--Painlev\'e structure considered in
Section~\ref{sec:toda-painleve}.
\section{Recursive computation and numerical safeguards}
\label{sec:recursive-computation}

The local Laguerre--Freud system derived in
Section~\ref{sec:laguerre-freud} provides a recursive construction of
the recurrence coefficients. Once the initial data have been
specified, the scheme generates the coefficients sequentially and
requires no moment matrix, determinant evaluation, or cumulative sums.

\subsection{Initialization}

The initial recurrence coefficients were obtained from the moments in
Section~\ref{sec:weight-moments}. They are
\begin{equation}
	\beta_0
	=
	\sqrt{\mu}\,
	\frac{I_1(2\sqrt{\mu})}
	{I_0(2\sqrt{\mu})},
	\label{eq:algorithm-beta-zero}
\end{equation}
and
\begin{equation}
	\gamma_1
	=
	\mu-\beta_0^2.
	\label{eq:algorithm-gamma-one}
\end{equation}

Applying \eqref{eq:laguerre-freud-first} at $n=1$ gives
\begin{equation}
	\beta_1
	=
	\frac{\mu}{\gamma_1}-\beta_0.
	\label{eq:algorithm-beta-one}
\end{equation}

Thus, the recursion is initialized by
\begin{equation}
	\beta_0,
	\qquad
	\gamma_1,
	\qquad
	\beta_1.
	\label{eq:algorithm-initial-data}
\end{equation}

\subsection{Local recursive scheme}

The fundamental relation governing the update of the subdiagonal
coefficient is the implicit Laguerre--Freud identity
\[
(\gamma_{n+1}-\mu)(\gamma_n-\mu)
=
\mu(\beta_n-n)^2,
\qquad
n\geq1,
\]
which is equation \eqref{eq:laguerre-freud-second}. This relation does
not require division by $\gamma_n-\mu$.

Whenever
\[
\gamma_n\neq\mu,
\]
the implicit relation may be solved explicitly for
$\gamma_{n+1}$:
\begin{equation}
	\gamma_{n+1}
	=
	\mu+
	\frac{
		\mu(\beta_n-n)^2
	}{
		\gamma_n-\mu
	},
	\qquad
	n\geq1.
	\label{eq:algorithm-gamma-recursion}
\end{equation}
Thus, \eqref{eq:algorithm-gamma-recursion} is a conditional explicit
update, rather than the fundamental Laguerre--Freud identity itself.

Once $\gamma_{n+1}$ has been computed, the first
Laguerre--Freud relation, applied at index $n+1$, yields
\begin{equation}
	\beta_{n+1}
	=
	\frac{\mu(n+1)}{\gamma_{n+1}}
	-\beta_n+n,
	\qquad
	n\geq1.
	\label{eq:algorithm-beta-recursion}
\end{equation}

Starting from \eqref{eq:algorithm-initial-data}, equations
\eqref{eq:algorithm-gamma-recursion} and
\eqref{eq:algorithm-beta-recursion} generate successively
\[
\gamma_2,\ \beta_2,\ \gamma_3,\ \beta_3,\ \ldots.
\]
Only the current pair $(\beta_n,\gamma_n)$ is needed at each step.
Consequently, the computation up to degree $N$ requires $O(N)$
arithmetic operations and $O(1)$ working memory. If the full sequence
of coefficients is retained for later use, the storage requirement is
$O(N)$.

\subsection{Numerical safeguards}

Since
\[
\gamma_n\longrightarrow\mu
\qquad
\text{as}
\qquad
n\longrightarrow\infty,
\]
the denominator
\[
\gamma_n-\mu
\]
in the explicit update \eqref{eq:algorithm-gamma-recursion} can become
small at large degree. The explicit recursive form may therefore lose
relative accuracy even though the implicit Laguerre--Freud identity
remains well defined. The computation should consequently be performed
using arbitrary-precision arithmetic whenever high-degree coefficients
are required.

A basic diagnostic is supplied by positivity. Since the generalized
Charlier measure is positive, the recurrence coefficients satisfy
\[
\gamma_n>0,
\qquad
n\geq1.
\]
A negative computed value of $\gamma_n$ is therefore incompatible with
the orthogonality measure and indicates loss of accuracy, incorrect
initialization, or an indexing error.

The local recursion should also be checked independently against the
moment/Hankel construction. The quantitative residual tests and the
comparison between the recursively generated coefficients and the
moment-based reference values are reported in
Section~\ref{sec:numerical-validation}.

\section{Toda deformation and the Painlev\'e V connection}
\label{sec:toda-painleve}

The local Laguerre--Freud system derived in
Section~\ref{sec:laguerre-freud} describes the dependence of the
recurrence coefficients on the polynomial degree $n$. We now regard
the weight parameter $\mu$ as a continuous deformation variable and
examine the corresponding differential relations.

The weight
\[
\rho_\mu(k)
=
\frac{\mu^k}{(k!)^2}
\]
is of exponential-deformation type. Consequently, its recurrence
coefficients satisfy the Toda equations. The Toda deformation provides
the continuous integrable structure associated with the discrete
Laguerre--Freud dynamics.

The Painlev\'e~V representations discussed below are known results for
the generalized Charlier family. They are recalled here in order to
connect the moment, Hankel, Laguerre--Freud, Toda, and numerical
descriptions of the recurrence coefficients within a common
normalization.

\subsection{Toda equations for the recurrence coefficients}

Let a prime denote differentiation with respect to $\mu$. The
recurrence coefficients satisfy the Toda system
\begin{equation}
	\mu\gamma_n'
	=
	\gamma_n
	\bigl(
	\beta_n-\beta_{n-1}
	\bigr),
	\qquad
	n\geq1,
	\label{eq:toda-gamma}
\end{equation}
and
\begin{equation}
	\mu\beta_n'
	=
	\gamma_{n+1}-\gamma_n,
	\qquad
	n\geq0.
	\label{eq:toda-beta}
\end{equation}

Equations \eqref{eq:toda-gamma} and \eqref{eq:toda-beta} should be
read together with the discrete Laguerre--Freud equations
\eqref{eq:laguerre-freud-first} and
\eqref{eq:laguerre-freud-second}. The discrete equations govern the
dependence on the degree $n$, whereas the Toda equations govern the
continuous dependence on the deformation parameter $\mu$.

In particular, equation \eqref{eq:toda-beta} gives
\begin{equation}
	\gamma_{n+1}
	=
	\gamma_n+\mu\beta_n'.
	\label{eq:gamma-from-toda-beta}
\end{equation}
Conversely, the first Laguerre--Freud relation gives
\begin{equation}
	\gamma_n
	=
	\frac{\mu n}
	{\beta_n+\beta_{n-1}-n+1},
	\qquad
	n\geq1.
	\label{eq:toda-gamma-from-beta}
\end{equation}

The combination of the Toda system with the Laguerre--Freud equations
leads to nonlinear differential equations for the recurrence
coefficients. Their reduction to Painlev\'e~V is known for the
generalized Charlier family; see \cite{FilipukVanAssche2013}.

\subsection{A Painlev\'e V representation of \texorpdfstring{$\beta_n$}{beta n}}

We use the standard fifth Painlev\'e equation
\begin{equation}
	y''
	=
	\left(
	\frac{1}{2y}
	+
	\frac{1}{y-1}
	\right)
	(y')^2
	-
	\frac{y'}{t}
	+
	\frac{(y-1)^2}{t^2}
	\left(
	Ay+\frac{B}{y}
	\right)
	+
	\frac{Cy}{t}
	+
	\frac{Dy(y+1)}{y-1},
	\label{eq:painleve-five-general}
\end{equation}
where the prime in this subsection denotes differentiation with respect
to $t$.

For the generalized Charlier specialization considered here, set
\begin{equation}
	t=\mu.
	\label{eq:painleve-first-variable}
\end{equation}
Then the diagonal recurrence coefficient $\beta_n(\mu)$ admits the
representation
\begin{equation}
	\beta_n(t)
	=
	\frac{
		n+1
		-
		(3n+1)y
		+
		2ny^2
		+
		ty'
	}{
		2y(y-1)
	},
	\label{eq:beta-from-painleve-first}
\end{equation}
where $y=y(t)$ satisfies \eqref{eq:painleve-five-general} with
parameters
\begin{equation}
	A=0,
	\qquad
	B=-\frac{(n+1)^2}{2},
	\qquad
	C=2,
	\qquad
	D=0.
	\label{eq:painleve-first-parameters}
\end{equation}

Equation \eqref{eq:beta-from-painleve-first} is the specialization to
the weight $\mu^k/(k!)^2$ of a known generalized Charlier
Painlev\'e~V representation. It does not constitute a new
Painlev\'e reduction.

\subsection{An equivalent Painlev\'e V parametrization}

For the same generalized Charlier weight, the recurrence coefficient
$\beta_n$ can also be represented in terms of a solution $y=y(z)$ of
\eqref{eq:painleve-five-general} with independent variable
\begin{equation}
	z^2=\mu
	\label{eq:painleve-second-variable}
\end{equation}
and parameters
\begin{equation}
	A=\frac{n^2}{8},
	\qquad
	B=-\frac{n^2}{8},
	\qquad
	C=0,
	\qquad
	D=-8.
	\label{eq:painleve-second-parameters}
\end{equation}

In this parametrization, one has
\begin{equation}
	\beta_n(\mu)
	=
	\frac{
		n
		+
		7ny^2
		-
		ny^3
		-
		2zy'
		-
		y(7n+2zy')
	}{
		8y(y-1)
	},
	\qquad
	\mu=z^2,
	\label{eq:beta-from-painleve-second}
\end{equation}
where the derivative in \eqref{eq:beta-from-painleve-second} is taken
with respect to $z$.

\subsection{Equivalence of the two Painlev\'e V representations}

The two parameter sets displayed above correspond to equivalent
Painlev\'e~V descriptions of the generalized Charlier recurrence
coefficients. Their apparent discrepancy is resolved by the
Painlev\'e~V transformations discussed explicitly in
\cite[Section~2.4]{FilipukVanAssche2013}. In particular, after
specialization to $\nu=1$, the two representations belong to the same
transformation framework and describe the same underlying generalized
Charlier recurrence-coefficient dynamics.

Since the explicit transformation chain is not required for the
developments that follow, we do not reproduce it here and refer the
reader to \cite{FilipukVanAssche2013} for its detailed construction.

It is important to emphasize that these Painlev\'e~V representations
are recalled here as part of the integrable structure of the
generalized Charlier family; no new Painlev\'e reduction is claimed in
the present work. Our purpose is to place them in a normalization
consistent with the weight
\[
\frac{\mu^k}{(k!)^2},
\]
and with the discrete Laguerre--Freud and Toda formulations used
throughout this paper.

\subsection{The subdiagonal coefficient}

The Painlev\'e~V representations above are stated for the diagonal
coefficient $\beta_n$. The coefficient $\gamma_n$ is then recovered
through either the discrete Laguerre--Freud relation
\eqref{eq:toda-gamma-from-beta} or the Toda equation
\eqref{eq:gamma-from-toda-beta}.

Thus, the differential reduction associated with $\gamma_n$ should not
be confused with the scalar Painlev\'e~V representation for
$\beta_n$. A separate scalar differential equation for $\gamma_n$ can be obtained
by eliminating $\beta_n$ from the corresponding Laguerre--Freud--Toda
differential system.  Since this equation is not
required for the recursive computation or the numerical validation, its
derivation is recorded in Appendix~\ref{app:gamma-ode}.

\section{Numerical cross-validation and asymptotic comparison}
\label{sec:numerical-validation}

\subsection{Moment-based reference computation}

The recurrence coefficients are computed independently from the moments
of the positive generalized Charlier measure
\begin{equation}
	\rho_\mu(k)
	=
	\frac{\mu^k}{(k!)^2},
	\qquad
	k\in\mathbb{N}_0,
	\qquad
	\mu>0.
	\label{eq:numerical-weight}
\end{equation}
The corresponding moments are
\begin{equation}
	M_r(\mu)
	=
	\sum_{k=0}^{\infty}
	k^r\frac{\mu^k}{(k!)^2},
	\qquad
	r\geq0.
	\label{eq:numerical-moments}
\end{equation}

For each fixed value of $\mu$, the moment series are evaluated by
adaptive truncation, with the truncation index chosen so that the
remaining tail is below the prescribed numerical tolerance. The monic
orthogonal polynomials are then constructed from the moments using
arbitrary-precision arithmetic. All computations reported in this
section use $100$ decimal digits of working precision.

The recurrence coefficients are recovered from the monic orthogonal
polynomials through
\begin{equation}
	\beta_n
	=
	\frac{\langle xP_n,P_n\rangle_\mu}
	{\langle P_n,P_n\rangle_\mu},
	\label{eq:numerical-beta}
\end{equation}
and
\begin{equation}
	\gamma_n
	=
	\frac{\langle P_n,P_n\rangle_\mu}
	{\langle P_{n-1},P_{n-1}\rangle_\mu},
	\qquad
	n\geq1.
	\label{eq:numerical-gamma}
\end{equation}
Since the orthogonality measure is positive,
\begin{equation}
	\gamma_n>0,
	\qquad
	n\geq1.
	\label{eq:numerical-gamma-positivity}
\end{equation}

The moment-based construction is independent of the local
Laguerre--Freud recursion and is therefore used as a reference method
for the numerical comparisons. The corresponding Hankel formulas and
implementation details are collected in
Appendix~\ref{app:moment-numerics}.

The parameter values considered below are
\begin{equation}
	\mu\in\{0.1,1,10\}.
	\label{eq:numerical-parameter-values}
\end{equation}

\subsection{Numerical recurrence coefficients}

Table~\ref{tab:first-ten-all-mu} displays the first recurrence
coefficients obtained from the moment-based construction. The numerical
values illustrate two basic features of the generalized Charlier
system.

First, the computed subdiagonal coefficients remain positive, in
agreement with \eqref{eq:numerical-gamma-positivity}. Second, the
transient regime becomes longer as $\mu$ increases. For $\mu=0.1$, the
coefficients are already close to their limiting values at small
degree. For $\mu=10$, the departure from the limiting regime remains
visible over a substantially larger range of degrees.

\begin{table}[H]
	\centering
	\caption{Moment-based recurrence coefficients $\beta_n$ and
		$\gamma_n$ for $\mu=0.1$, $\mu=1$, and $\mu=10$.}
	\label{tab:first-ten-all-mu}
	\renewcommand{\arraystretch}{1.15}
	\resizebox{\textwidth}{!}{%
		\begin{tabular}{rrrrrrr}
			\toprule
			&
			\multicolumn{2}{c}{$\mu=0.1$}
			&
			\multicolumn{2}{c}{$\mu=1$}
			&
			\multicolumn{2}{c}{$\mu=10$}
			\\
			\cmidrule(lr){2-3}
			\cmidrule(lr){4-5}
			\cmidrule(lr){6-7}
			$n$
			&
			$\beta_n$
			&
			$\gamma_n$
			&
			$\beta_n$
			&
			$\gamma_n$
			&
			$\beta_n$
			&
			$\gamma_n$
			\\
			\midrule
			$0$
			&
			$0.0953118976$
			&
			--
			&
			$0.6977746580$
			&
			--
			&
			$2.9002024851$
			&
			--
			\\
			$1$
			&
			$1.0046088378$
			&
			$0.0909156422$
			&
			$1.2511231876$
			&
			$0.5131105267$
			&
			$3.3937547269$
			&
			$1.5888255454$
			\\
			$2$
			&
			$2.0000785975$
			&
			$0.0997661763$
			&
			$2.0464648008$
			&
			$0.8704780884$
			&
			$3.8806359466$
			&
			$3.1875605204$
			\\
			$3$
			&
			$3.0000006638$
			&
			$0.0999973580$
			&
			$3.0043893444$
			&
			$0.9833311778$
			&
			$4.3585322921$
			&
			$4.8083332349$
			\\
			$4$
			&
			$4.0000000034$
			&
			$0.0999999833$
			&
			$4.0002393303$
			&
			$0.9988441688$
			&
			$4.8477781347$
			&
			$6.4450530588$
			\\
			$5$
			&
			$5.0000000000$
			&
			$0.0999999999$
			&
			$5.0000084647$
			&
			$0.9999504435$
			&
			$5.4192744469$
			&
			$7.9782320873$
			\\
			$6$
			&
			$6.0000000000$
			&
			$0.1000000000$
			&
			$6.0000002104$
			&
			$0.9999985542$
			&
			$6.1521012833$
			&
			$9.1305082015$
			\\
			$7$
			&
			$7.0000000000$
			&
			$0.1000000000$
			&
			$7.0000000039$
			&
			$0.9999999694$
			&
			$7.0392406717$
			&
			$9.7339273306$
			\\
			$8$
			&
			$8.0000000000$
			&
			$0.1000000000$
			&
			$8.0000000001$
			&
			$0.9999999995$
			&
			$8.0073268644$
			&
			$9.9421274526$
			\\
			$9$
			&
			$9.0000000000$
			&
			$0.1000000000$
			&
			$9.0000000000$
			&
			$1.0000000000$
			&
			$9.0010293440$
			&
			$9.9907239366$
			\\
			\bottomrule
		\end{tabular}%
	}
\end{table}

\begin{figure}[H]
	\centering
	\includegraphics[width=0.82\textwidth]{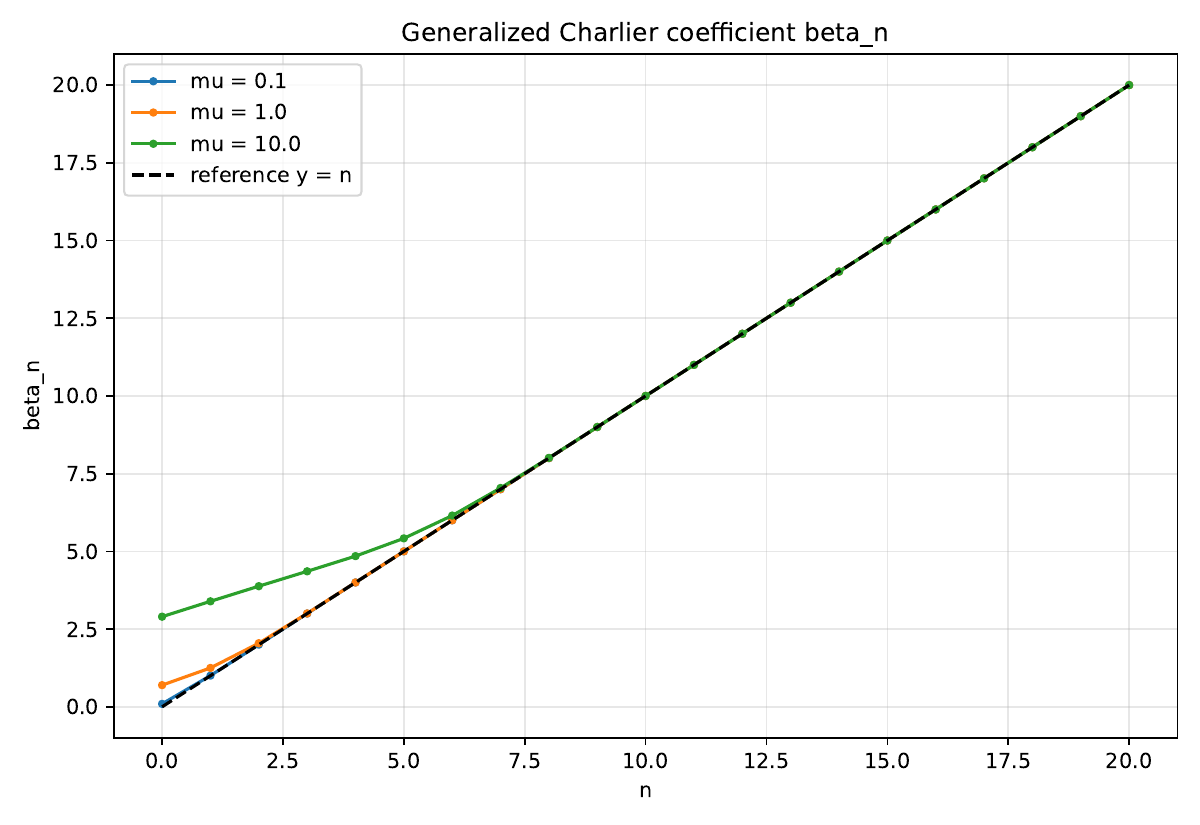}
	\caption{Moment-based recurrence coefficient $\beta_n$ as a function
		of $n$ for $\mu=0.1$, $\mu=1$, and $\mu=10$. The dashed line is the
		reference line $y=n$.}
	\label{fig:beta-coefficients}
\end{figure}

Figure~\ref{fig:beta-coefficients} gives a graphical representation of
the diagonal recurrence coefficient $\beta_n$. The dashed reference
line corresponds to the leading term $n$.

\begin{figure}[H]
	\centering
	\includegraphics[width=0.82\textwidth]{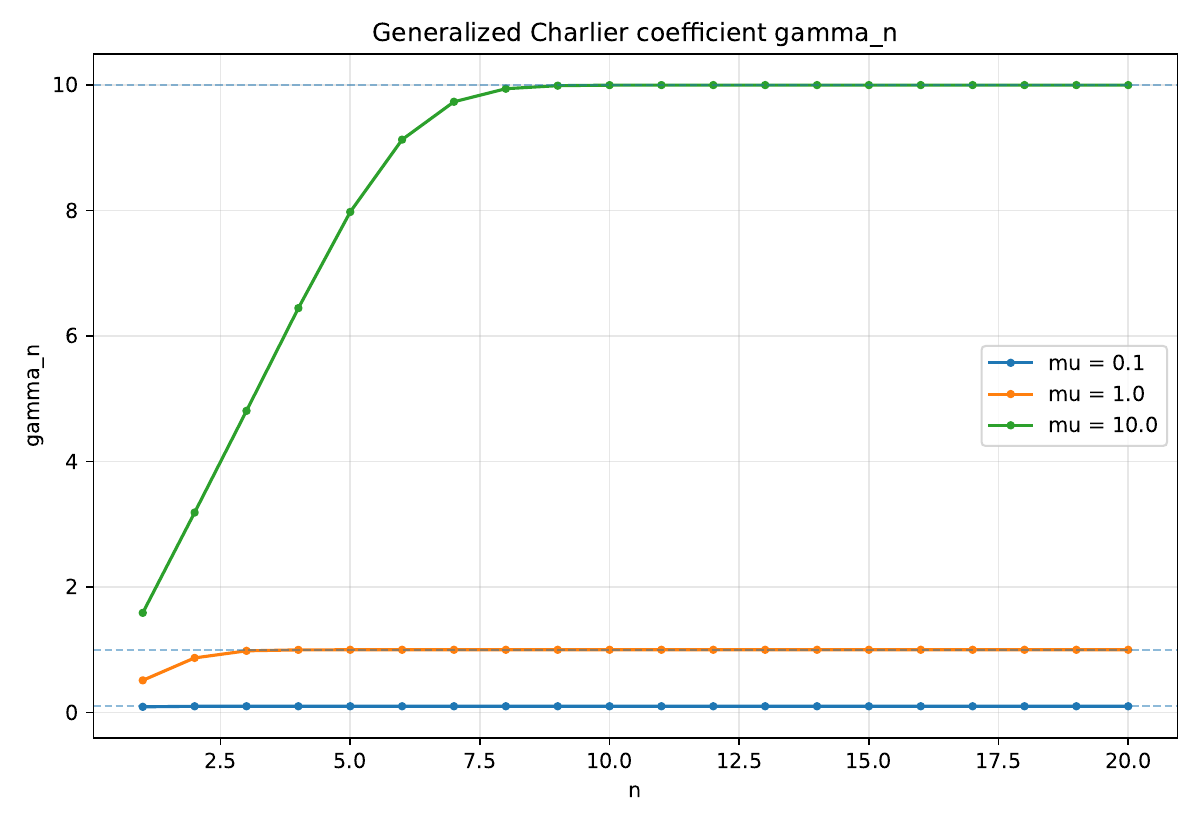}
	\caption{Moment-based recurrence coefficient $\gamma_n$ as a function
		of $n$ for $\mu=0.1$, $\mu=1$, and $\mu=10$. The horizontal dashed
		lines indicate the corresponding levels $y=\mu$.}
	\label{fig:gamma-coefficients}
\end{figure}

Figure~\ref{fig:gamma-coefficients} displays the corresponding
subdiagonal coefficient $\gamma_n$. The horizontal dashed lines
indicate the levels $y=\mu$.

Together, Figures~\ref{fig:beta-coefficients} and
\ref{fig:gamma-coefficients} illustrate the approach of $\beta_n$ to
$n$ and of $\gamma_n$ to $\mu$. They also show that the transient
regime is longer for larger values of $\mu$.

\subsection{Cross-validation of the discrete system}

The moment-based coefficients can be tested against the specialized
discrete Laguerre--Freud system
\begin{equation}
	\beta_n+\beta_{n-1}-n+1
	=
	\frac{\mu n}{\gamma_n},
	\qquad
	n\geq1,
	\label{eq:numerical-discrete-system-first}
\end{equation}
and
\begin{equation}
	(\gamma_{n+1}-\mu)(\gamma_n-\mu)
	=
	\mu(\beta_n-n)^2.
	\label{eq:numerical-discrete-system-second}
\end{equation}

The corresponding residuals are
\begin{equation}
	R_n^{(1)}
	=
	\beta_n+\beta_{n-1}-n+1
	-
	\frac{\mu n}{\gamma_n},
	\qquad
	n\geq1,
	\label{eq:numerical-residual-one}
\end{equation}
and
\begin{equation}
	R_n^{(2)}
	=
	(\gamma_{n+1}-\mu)(\gamma_n-\mu)
	-
	\mu(\beta_n-n)^2.
	\label{eq:numerical-residual-two}
\end{equation}

For a chosen degree cutoff $N$, we report
\begin{equation}
	\max_{1\leq n\leq N}
	|R_n^{(1)}|,
	\qquad
	\max_{1\leq n\leq N-1}
	|R_n^{(2)}|,
	\label{eq:numerical-residual-maxima}
\end{equation}
together with the differences between the recurrence coefficients
generated by the local Laguerre--Freud recursion and those obtained
independently from the moment/Hankel construction:
\begin{equation}
	\max_{0\leq n\leq N}
	\left|
	\beta_n^{\mathrm{LF}}-\beta_n^{\mathrm{H}}
	\right|,
	\qquad
	\max_{1\leq n\leq N}
	\left|
	\gamma_n^{\mathrm{LF}}-\gamma_n^{\mathrm{H}}
	\right|.
	\label{eq:numerical-cross-validation-errors}
\end{equation}

Here the superscripts $\mathrm{LF}$ and $\mathrm{H}$ denote,
respectively, the coefficients generated by the local
Laguerre--Freud recursion and those computed from the independent
moment/Hankel construction.

\begin{table}[H]
	\centering
	\caption{Numerical cross-validation of the local Laguerre--Freud
		recursion against the independent moment/Hankel construction. The
		computations use $100$ decimal digits of working precision and
		$N=20$.}
	\label{tab:numerical-cross-validation}
	\renewcommand{\arraystretch}{1.15}
	\resizebox{\textwidth}{!}{%
		\begin{tabular}{ccrrrr}
			\toprule
			$\mu$
			&
			$N$
			&
			$\displaystyle\max_{1\leq n\leq N}|R_n^{(1)}|$
			&
			$\displaystyle\max_{1\leq n\leq N-1}|R_n^{(2)}|$
			&
			$\displaystyle\max_{0\leq n\leq N}
			|\beta_n^{\mathrm{LF}}-\beta_n^{\mathrm{H}}|$
			&
			$\displaystyle\max_{1\leq n\leq N}
			|\gamma_n^{\mathrm{LF}}-\gamma_n^{\mathrm{H}}|$
			\\
			\midrule
			$0.1$
			&
			$20$
			&
			$2.29\times10^{-100}$
			&
			$2.40\times10^{-105}$
			&
			$2.83\times10^{-43}$
			&
			$1.31\times10^{-45}$
			\\
			$1$
			&
			$20$
			&
			$2.29\times10^{-100}$
			&
			$8.93\times10^{-103}$
			&
			$6.08\times10^{-60}$
			&
			$1.37\times10^{-61}$
			\\
			$10$
			&
			$20$
			&
			$2.29\times10^{-100}$
			&
			$1.71\times10^{-100}$
			&
			$4.54\times10^{-72}$
			&
			$1.13\times10^{-72}$
			\\
			\bottomrule
		\end{tabular}%
	}
\end{table}

Table~\ref{tab:numerical-cross-validation} provides a quantitative
comparison between the local Laguerre--Freud recursion and the
independent moment/Hankel construction. The residuals of the discrete
system are close to the working precision, confirming the internal
consistency of the recursively generated coefficients.

The comparison with the moment/Hankel computation is less accurate
than the internal residual test, particularly for smaller values of
$\mu$. This is consistent with increased numerical sensitivity as
$\gamma_n$ approaches its limiting value $\mu$, since the explicit
local update contains the small denominator
\[
\gamma_n-\mu.
\]
Conditioning effects in the moment/Hankel construction may also
contribute to the observed discrepancy. Nevertheless, the two
independently constructed coefficient sequences agree to many decimal
digits for all parameter values and degrees considered here.
%

\subsection{Comparison with the leading asymptotics}

The numerical coefficients are consistent with the leading
large-degree relations
\begin{equation}
	\beta_n\sim n,
	\qquad
	\gamma_n\sim\mu,
	\qquad
	n\to\infty.
	\label{eq:numerical-leading-asymptotics}
\end{equation}

For the generalized Charlier weight considered here, these leading
relations are consistent with the known asymptotic behavior of the
recurrence coefficients. In particular, the diagonal coefficient grows
linearly with $n$, whereas the subdiagonal coefficient approaches the
finite limiting value $\mu$.

The numerical values in Table~\ref{tab:first-ten-all-mu} and the
coefficient plots in Figures~\ref{fig:beta-coefficients} and
\ref{fig:gamma-coefficients} are consistent with
\eqref{eq:numerical-leading-asymptotics}. Determining higher-order
corrections in the specialization considered here lies beyond the
scope of the present work.

\section{Conclusion}
\label{sec:conclusion}

We have studied the monic orthogonal polynomials associated with the
generalized Charlier weight
\[
\rho_\mu(k)
=
\frac{\mu^k}{(k!)^2},
\qquad
k\in\mathbb{N}_0,
\qquad
\mu>0,
\]
together with their recurrence coefficients $\beta_n$ and
$\gamma_n$. This weight is the specialization
\[
a=\mu,
\qquad
\nu=1
\]
of the two-parameter generalized Charlier family
\[
w_k(a,\nu)
=
\frac{a^k}{(\nu)_k k!}.
\]

The principal purpose of this work has been to give a consistent
specialized account of several complementary descriptions of the
recurrence coefficients. At the moment level, the coefficients are
characterized by Hankel determinants. The subdiagonal coefficient is
given by
\[
\gamma_n
=
\frac{\Delta_{n+1}\Delta_{n-1}}{\Delta_n^2},
\qquad
n\geq1,
\]
whereas the diagonal coefficient is expressed through the difference
of two ratios involving bordered Hankel determinants. The latter
distinction is essential: the fully shifted Hankel determinant is not
the determinant entering the formula for $\beta_n$.

The corresponding Markov-function description requires an equally
important normalization. The unnormalized Markov function has leading
behavior
\[
m(z)
=
\frac{M_0(\mu)}{z}
+
O(z^{-2}),
\qquad
z\to\infty.
\]
Consequently, the Jacobi continued fraction with numerator $1$
belongs to the normalized function
\[
\widehat m(z)
=
\frac{m(z)}{M_0(\mu)}.
\]
Its Jacobi parameters are precisely the recurrence coefficients
$\beta_n$ and $\gamma_n$.

We have also clarified the relation between two forms of the
Laguerre--Freud dynamics. Starting from the original
Hounkonnou--Hounga--Ronveaux equations, which contain cumulative sums
of recurrence coefficients, we introduced the shifted coefficient
\[
b_n
=
\beta_n-n.
\]
The cumulative quantities can then be eliminated, yielding the local
system
\[
\beta_n+\beta_{n-1}-n+1
=
\frac{\mu n}{\gamma_n},
\]
and
\[
(\gamma_{n+1}-\mu)(\gamma_n-\mu)
=
\mu(\beta_n-n)^2.
\]
Thus, the local recursive formulation is not a competing
Laguerre--Freud system: it is a reduction of the original
Hounkonnou--Hounga--Ronveaux formulation in the present generalized
Charlier specialization.

When the parameter $\mu$ is treated as a continuous deformation
variable, the recurrence coefficients satisfy the Toda equations. The
combination of the Toda and Laguerre--Freud structures leads to the
known Painlev\'e~V representations for the generalized Charlier
family. In this paper, these representations have been recalled in a
normalization consistent with the weight $\mu^k/(k!)^2$. No new
Painlev\'e reduction is claimed. The scalar differential relation for
$\gamma_n$ is distinct from the Painlev\'e~V representation of
$\beta_n$ and has been treated separately. In the resulting
Painlevé~V equation, the parameters depend explicitly on $n$, while
$t=\mu$ plays the role of the continuous deformation variable.

The moment/Hankel construction and moment-based orthogonalization
provide complementary moment-theoretic realizations of the recurrence
coefficients, whereas the local Laguerre--Freud recursion supplies an
independent dynamical construction. Their comparison therefore
provides effective numerical diagnostic checks for normalization,
indexing, and numerical sensitivity in the recursive scheme. The
numerical results confirm the positivity of $\gamma_n$, illustrate the
dependence of the transient regime on $\mu$, and are consistent with
the leading large-degree relations
\[
\beta_n\sim n,
\qquad
\gamma_n\sim\mu,
\qquad
n\to\infty.
\]

It is instructive to compare this behavior with that of the classical
Charlier weight
\[
w_{\mathrm{Ch}}(k)
=
\frac{a^k}{k!}.
\]
For the classical Charlier family, the monic recurrence coefficients
are explicit:
\[
\beta_n^{\mathrm{Ch}}
=
n+a,
\qquad
\gamma_n^{\mathrm{Ch}}
=
na.
\]
For the generalized weight studied here, the leading behavior is
instead
\[
\beta_n\sim n,
\qquad
\gamma_n\sim\mu.
\]
The additional factorial factor in $\mu^k/(k!)^2$ therefore changes
the recurrence structure qualitatively: the subdiagonal coefficient,
which grows linearly in the classical Charlier case, remains bounded
and tends to the finite limit $\mu$.

Taken together, the moment/Hankel, Markov--Jacobi, Laguerre--Freud,
and Toda--Painlevé viewpoints provide a coherent framework that
realizes, for the specialization $\nu=1$, the four mutually consistent
and complementary descriptions stated informally in the introduction.
Beyond the specific case of the generalized Charlier weight, this
unified approach can serve as a template for organizing the various
descriptions of other discrete semiclassical families.

Determining higher-order asymptotic corrections to
$\beta_n\sim n$ and $\gamma_n\sim\mu$, together with rigorous and
uniform remainder estimates, remains an interesting asymptotic
problem. Further work may also address the numerical stability of the
local Laguerre--Freud recursion in the regime where
$\gamma_n-\mu$ is small. A systematic comparison among recursive
iteration, moment-based orthogonalization, and Hankel-determinant
computations could yield quantitative error estimates and clarify the
range of degrees for which each construction is most effective.
Extensions to other generalized Charlier weights and related discrete
semiclassical families offer further directions for studying the
interaction among moment representations, recurrence-coefficient
dynamics, and integrable structures.
\section*{Appendices}
\appendix

\section{A scalar differential equation for \texorpdfstring{$\gamma_n$}{gamma n}}
\label{app:gamma-ode}

This appendix records a scalar differential reduction for the
subdiagonal recurrence coefficient $\gamma_n$. It is distinct from the
Painlev\'e~V representation of the diagonal recurrence coefficient
$\beta_n$ discussed in Section~\ref{sec:toda-painleve}.


In Section~\ref{sec:toda-painleve}, the diagonal coefficient is
described through a known Painlev\'e~V representation. Here, we use
the Laguerre--Freud--Toda formulation for the generalized Charlier
family developed in \cite{FernandezIrisarriManas2023}. Starting from
the corresponding first-order differential system for
$(\beta_n,\gamma_n)$, we eliminate $\beta_n$ to obtain a scalar
second-order nonlinear differential equation for $\gamma_n$.

The notation in this appendix is independent of the shifted
coefficient $b_n$ introduced in Section~\ref{sec:laguerre-freud},
where
\[
b_n=\beta_n-n.
\]
To avoid a conflict with that notation, the parameter of the general
generalized Charlier weight is denoted below by $\alpha$.

Consider the weight
\begin{equation}
	w(k)
	=
	\frac{\eta^k}{(\alpha+1)_k k!},
	\qquad
	k\in\mathbb{N}_0,
	\qquad
	\eta>0,
	\qquad
	\alpha>-1.
	\label{eq:app-generalized-charlier-weight}
\end{equation}
The weight studied in the present paper is recovered by setting
\begin{equation}
	\alpha=0,
	\qquad
	\eta=\mu.
	\label{eq:app-nu-one-specialization}
\end{equation}

Let
\begin{equation}
	\vartheta_\eta
	=
	\eta\frac{d}{d\eta}
	\label{eq:app-euler-operator}
\end{equation}
denote the Euler differential operator. The monic recurrence relation
is
\begin{equation}
	xP_n(x)
	=
	P_{n+1}(x)
	+
	\beta_nP_n(x)
	+
	\gamma_nP_{n-1}(x).
	\label{eq:app-monic-recurrence}
\end{equation}

For the generalized Charlier weight
\eqref{eq:app-generalized-charlier-weight}, the relevant
Laguerre--Freud--Toda equations take the form
\begin{equation}
	\vartheta_\eta\beta_n
	=
	\frac{
		\eta^2+(\alpha-n)n+(2n-\alpha-\beta_n)\beta_n-\gamma_n
	}{
		\eta-\gamma_n
	}
	-\gamma_n,
	\label{eq:app-beta-first-order-system}
\end{equation}
and
\begin{equation}
	\vartheta_\eta\gamma_n
	=
	(\alpha-n+1+2\beta_n)\gamma_n-n\eta.
	\label{eq:app-gamma-first-order-system}
\end{equation}

Equation \eqref{eq:app-gamma-first-order-system} is linear in
$\beta_n$. Therefore,
\begin{equation}
	\beta_n
	=
	\frac{1}{2}
	\left(
	\frac{\vartheta_\eta\gamma_n}{\gamma_n}
	+
	\frac{n\eta}{\gamma_n}
	-\alpha+n-1
	\right).
	\label{eq:app-beta-from-gamma}
\end{equation}

Define
\begin{equation}
	A_n
	=
	\frac{\vartheta_\eta\gamma_n}{\gamma_n}
	+
	\frac{n\eta}{\gamma_n}.
	\label{eq:app-An-definition}
\end{equation}
Then
\begin{equation}
	\beta_n
	=
	\frac{1}{2}(A_n-\alpha+n-1),
	\qquad
	\vartheta_\eta\beta_n
	=
	\frac{1}{2}\vartheta_\eta A_n.
	\label{eq:app-beta-An}
\end{equation}

Substitution of \eqref{eq:app-beta-An} into
\eqref{eq:app-beta-first-order-system} yields the scalar second-order
equation
\begin{align}
	&
	\left(
	1-\frac{\gamma_n}{\eta}
	\right)
	\left[
	\vartheta_\eta
	\left(
	\frac{\vartheta_\eta\gamma_n}{\gamma_n}
	+
	\frac{n\eta}{\gamma_n}
	\right)
	+
	2\gamma_n
	\right]
	+
	2\left[
	\gamma_n-\eta+(n-\alpha)n
	\right]
	\nonumber\\
	&\qquad
	=
	-\frac{1}{2}
	\left(
	\frac{\vartheta_\eta\gamma_n}{\gamma_n}
	+
	\frac{n\eta}{\gamma_n}
	\right)^2
	+
	(n+1)
	\left(
	\frac{\vartheta_\eta\gamma_n}{\gamma_n}
	+
	\frac{n\eta}{\gamma_n}
	\right)
	\nonumber\\
	&\qquad\quad
	+
	(n-\alpha-1)(3n-\alpha+1).
	\label{eq:app-gamma-second-order-general}
\end{align}

Equation \eqref{eq:app-gamma-second-order-general} is a scalar
second-order nonlinear differential equation for $\gamma_n(\eta)$.
It is obtained by eliminating $\beta_n$ from the
Laguerre--Freud--Toda system and constitutes a distinct reduction from
the Painlev\'e~V representation stated for $\beta_n$ in
Section~\ref{sec:toda-painleve}.

\paragraph{Specialization to
	\texorpdfstring{$\mu^k/(k!)^2$}{mu^k/(k!)^2}.}

For the weight
\begin{equation}
	\rho_\mu(k)
	=
	\frac{\mu^k}{(k!)^2},
	\label{eq:app-specialized-weight}
\end{equation}
we set
\begin{equation}
	\alpha=0,
	\qquad
	\eta=\mu,
	\qquad
	\vartheta_\mu
	=
	\mu\frac{d}{d\mu}.
	\label{eq:app-specialized-variables}
\end{equation}
Equation \eqref{eq:app-gamma-second-order-general} becomes
\begin{align}
	&
	\left(
	1-\frac{\gamma_n}{\mu}
	\right)
	\left[
	\vartheta_\mu
	\left(
	\frac{\vartheta_\mu\gamma_n}{\gamma_n}
	+
	\frac{n\mu}{\gamma_n}
	\right)
	+
	2\gamma_n
	\right]
	+
	2(\gamma_n-\mu+n^2)
	\nonumber\\
	&\qquad
	=
	-\frac{1}{2}
	\left(
	\frac{\vartheta_\mu\gamma_n}{\gamma_n}
	+
	\frac{n\mu}{\gamma_n}
	\right)^2
	+
	(n+1)
	\left(
	\frac{\vartheta_\mu\gamma_n}{\gamma_n}
	+
	\frac{n\mu}{\gamma_n}
	\right)
	+
	(n-1)(3n+1).
	\label{eq:app-gamma-second-order-specialized}
\end{align}

The diagonal recurrence coefficient is reconstructed from
$\gamma_n$ by
\begin{equation}
	\beta_n
	=
	\frac{1}{2}
	\left(
	\frac{\mu\gamma_n'(\mu)}{\gamma_n(\mu)}
	+
	\frac{n\mu}{\gamma_n(\mu)}
	+n-1
	\right),
	\label{eq:app-beta-from-gamma-specialized}
\end{equation}
where
\begin{equation}
	\gamma_n'(\mu)
	=
	\frac{d\gamma_n}{d\mu}.
	\label{eq:app-gamma-prime}
\end{equation}

For reference, if
\begin{equation}
	g(\mu)
	=
	\gamma_n(\mu),
	\label{eq:app-g-definition}
\end{equation}
then
\begin{equation}
	\vartheta_\mu
	\left(
	\frac{\vartheta_\mu g}{g}
	+
	\frac{n\mu}{g}
	\right)
	=
	\frac{
		\mu g'+\mu^2g''+n\mu
	}{
		g
	}
	-
	\frac{
		\mu^2g'(g'+n)
	}{
		g^2
	}.
	\label{eq:app-expanded-euler-term}
\end{equation}
Consequently, \eqref{eq:app-gamma-second-order-specialized} is an
explicit second-order ordinary differential equation involving only
$g(\mu)$, $g'(\mu)$, $g''(\mu)$, $\mu$, and $n$.

This scalar equation is included as a supplementary continuous
reduction of the generalized Charlier recurrence-coefficient dynamics.
It is not used in the local recursive algorithm of
Section~\ref{sec:recursive-computation} or in the numerical
cross-validation of Section~\ref{sec:numerical-validation}. Its role
is to make explicit that the differential reduction for $\gamma_n$ is
distinct from the Painlev\'e~V representation stated for $\beta_n$.

\section{Details of the moment-based computation}
\label{app:moment-numerics}

This appendix summarizes the numerical procedure used to construct an
independent moment-based reference sequence for the recurrence
coefficients. The method is used in
Section~\ref{sec:numerical-validation} to validate the local
Laguerre--Freud recursion.

For each fixed value of $\mu$, the moments are evaluated from
\begin{equation}
	M_r(\mu)
	=
	\sum_{k=0}^{\infty}
	k^r\frac{\mu^k}{(k!)^2},
	\qquad
	r\geq0.
	\label{eq:app-moment-series}
\end{equation}
The series are truncated adaptively. More precisely, the truncation
index is increased until the estimated contribution of the remaining
tail is below the prescribed working tolerance. All computations
reported in Section~\ref{sec:numerical-validation} use $100$ decimal
digits of working precision.

The terms in the moment series are generated recursively. If
\begin{equation}
	t_k
	=
	\frac{\mu^k}{(k!)^2},
	\label{eq:app-moment-term}
\end{equation}
then
\begin{equation}
	t_0=1,
	\qquad
	t_{k+1}
	=
	\frac{\mu}{(k+1)^2}t_k.
	\label{eq:app-moment-term-recurrence}
\end{equation}
Thus, the moments are accumulated from
\begin{equation}
	M_r(\mu)
	=
	\sum_{k=0}^{\infty}k^r t_k.
	\label{eq:app-moment-accumulation}
\end{equation}

Once sufficiently many moments have been computed, the monic
orthogonal polynomials may be obtained by arbitrary-precision
Gram--Schmidt orthogonalization of
\[
1,\ x,\ x^2,\ \ldots,\ x^N
\]
with respect to the moment functional. Equivalently, one may use a
Cholesky factorization of the truncated Hankel moment matrix
\begin{equation}
	H_N
	=
	[M_{j+k}]_{j,k=0}^{N}.
	\label{eq:app-moment-matrix}
\end{equation}

The recurrence coefficients are then recovered from
\begin{equation}
	\beta_n
	=
	\frac{\langle xP_n,P_n\rangle_\mu}
	{\langle P_n,P_n\rangle_\mu},
	\label{eq:app-moment-beta}
\end{equation}
and
\begin{equation}
	\gamma_n
	=
	\frac{\langle P_n,P_n\rangle_\mu}
	{\langle P_{n-1},P_{n-1}\rangle_\mu},
	\qquad
	n\geq1.
	\label{eq:app-moment-gamma}
\end{equation}

The same coefficients can alternatively be checked using the
Hankel-determinant identities of
Section~\ref{sec:hankel-coefficients}. Together, these
moment-theoretic constructions provide a reference comparison
independent of the local Laguerre--Freud recursion.

For a degree cutoff $N$, the diagnostic quantities used in
Section~\ref{sec:numerical-validation} are the residual maxima
\[
\max_{1\leq n\leq N}|R_n^{(1)}|,
\qquad
\max_{1\leq n\leq N-1}|R_n^{(2)}|,
\]
together with
\[
\max_{0\leq n\leq N}
\left|
\beta_n^{\mathrm{LF}}-\beta_n^{\mathrm{H}}
\right|,
\qquad
\max_{1\leq n\leq N}
\left|
\gamma_n^{\mathrm{LF}}-\gamma_n^{\mathrm{H}}
\right|.
\]
The residuals $R_n^{(1)}$ and $R_n^{(2)}$ are defined in
Section~\ref{sec:numerical-validation}.

\section*{Conflict of interest}

The author declares that there is no conflict of interest regarding the
publication of this article.

\end{document}